\documentclass[journal]{IEEEtran}
\usepackage{cite}
\usepackage{amsmath,amssymb,amsfonts}
\usepackage{algorithmic}
\usepackage{graphicx}
\usepackage{textcomp}
\usepackage{booktabs}
\usepackage{multirow}
\usepackage{url}
\usepackage{float}
\usepackage[colorlinks=true,allcolors=blue]{hyperref}
\def\BibTeX{{\rm B\kern-.05em{\sc i\kern-.025em b}\kern-.08em
    T\kern-.1667em\lower.7ex\hbox{E}\kern-.125emX}}
\begin{document}

\title{A Sim-to-Real Study of Surface-Code Decoder Benchmarking}

\author{Shay~J.~Manor,~Leila~S.~Erhili,~and~Yassine~Jebbouri%
\thanks{S. J. Manor is with the Department of Computer Science, Purdue University, West Lafayette, IN 47907 USA (e-mail: manors@purdue.edu).}%
\thanks{L. S. Erhili is with the Department of Computer Science, Columbia University, New York, NY 10027 USA.}%
\thanks{Y. Jebbouri is with the Data Science Institute, Columbia University, New York, NY 10027 USA.}%
\thanks{This work has been submitted to the IEEE for possible publication. Copyright may be transferred without notice, after which this version may no longer be accessible.}
\thanks{\textit{Data and code availability:} All data and methods used in this study are publicly available. The device data is the public below-threshold memory-experiment release accompanying~\cite{willow2024}; no other source of hardware data was used. The full pipeline, the four noise-model implementations, the decoder harness, and the per-shot outcome of every evaluation reported here are released at \url{https://github.com/ShayManor/Sim2Real-Decoder-Benchmark}.}%
\thanks{\textit{Disclaimer:} This work was conducted independently, outside the scope of any employment, and using no internal, proprietary, or otherwise non-public data, code, models, or computing resources of any company. All hardware data used here is the public release cited as~\cite{willow2024}. The work was not funded, directed, reviewed, or endorsed by any company, and the views, analyses, and conclusions are those of the authors alone. They do not represent the positions of any organization with which the authors are or have been affiliated.}}

\markboth{Preprint}%
{Manor \MakeLowercase{\textit{et al.}}: A Sim-to-Real Study of Surface-Code Decoder Benchmarking}


\maketitle

\begin{abstract}
Quantum error-correction decoders are typically benchmarked against synthetic circuit-level noise, under the assumption that a decoder's ranking under such noise transfers to hardware and improves as the noise model becomes more realistic. The Willow processor, the first to operate below the surface-code threshold, allows us to test this assumption. We rank a panel of six decoders using a four-rung ladder of noise models with increasing fidelity, evaluated against real data across three code distances, two bases, and fifteen round counts. Rank agreement with hardware appears once the noise model gives each operation type its own error rate. Calibrating the model to the device improves absolute error rates but not rank agreement. We additionally provide the first independent evaluation of NVIDIA's Ising pre-decoder on hardware, at code distances below its training receptive field and via a mapping onto the lattice on which it was trained. Under these conditions, it holds no accuracy-latency advantage: another panel decoder matches or improves on it in both per-cycle error rate and decode latency in 278 of the 280 evaluations. We release the full pipeline and the per-shot outcome of every evaluation, so future decoders and devices can be compared.
\end{abstract}
\begin{IEEEkeywords}
Benchmark testing, decoding, error correction codes, logical error
rate, machine learning, noise measurement, quantum computing, quantum
error correction, surface code.
\end{IEEEkeywords}

\IEEEpeerreviewmaketitle

\section{Introduction}
\label{sec:introduction}


Quantum computers can solve problems beyond classical reach, but their
physical qubits are noisy, and the errors accumulate until a computation
fails.
Quantum error correction (QEC) keeps this noise in check, encoding a logical qubit across many physical qubits and using a
decoder to infer, from the observed syndrome, what errors occurred
and how to reverse them. The decoder is a part of this process that can be improved without changing the hardware:
for a fixed device, a better decoder means a lower logical error rate, which
translates directly into how deep a computation the machine can run before
noise overwhelms it. Choosing the right decoder therefore matters, and
choosing it correctly depends on being able to benchmark decoders in
a way that reflects how they will actually perform on hardware.

New decoders are validated
almost exclusively against synthetic circuit-level noise models~\cite{higgott2023beliefmatching,beni2025tesseract}: idealized
or partially calibrated stochastic Pauli channels standing in for a
device's error process, which also produces leakage, crosstalk, drift, soft
readout, and cosmic-ray
bursts~\cite{leakage2021,sarovar2020crosstalk,klimov2018drift,pattison2021soft,cosmicray2022}.
NVIDIA's Ising pre-decoder for surface
codes~\cite{nvidia_ising2026} is representative: it is designed to improve the accuracy--latency frontier
of an existing matching decoder, and its reported gains were established entirely
in simulation, benchmarked against the same pipeline without the pre-decoder. Benchmarking this way assumes a
synthetic ranking transfers to real hardware, and transfers better as the
noise model becomes more realistic.

The public release of below-threshold memory experiments from the
Willow processor~\cite{willow2024} lets us test this assumption. Earlier devices also released multi-distance
syndrome data~\cite{google2023scaling}; Willow is the first to run below the
surface-code threshold, at distances up to 7. On this data the assumption
fails in two ways: rank agreement between a synthetic model and hardware does
not rise monotonically with noise-model fidelity, and it varies with code
distance and round count. The same two decoders can rank correctly at one code distance and swap at
another, even under a better-calibrated model. A ranking computed under a
single noise model can therefore mislead, however realistic that model is. In this paper we use the Willow
dataset as ground truth and a graded ladder of noise models to test where
synthetic-model rankings hold and where they break down, including whether
NVIDIA's pre-decoder retains its reported advantage once evaluated under
these conditions.
\subsection{Contributions}
\begin{enumerate}
    \item \textbf{A noise-fidelity ladder.} We introduce a controlled ablation
    that measures whether a synthetic benchmark ranks decoders in the same way as the hardware by decoding identical circuits under noise models of increasing
    fidelity.

    \item \textbf{A hardware-grounded benchmark.} We establish the decoder
    ranking on a below-threshold device and measure how well each synthetic
    noise model predicts it (Section~\ref{sec:rank_ladder}).

    \item \textbf{An independent evaluation of the Ising pre-decoder.} We give
    the first assessment of NVIDIA's pre-decoder by anyone outside its authors,
    and find no advantage on hardware.

    \item \textbf{An open pipeline.} We release the pipeline and per-shot
    results so that decoder authors without hardware access can check their
    synthetic claims against a device.
\end{enumerate}

\section{Background and Related Work}

\subsection{The Decoder Landscape}

No decoder is uniformly best: the choice depends on the code family, the
noise model, and the latency
budget~\cite{dennis2002topological,fowler2012surfacecodes}. Decoder families
therefore range from fast matching to slower, more accurate most-likely-error
and learned decoders. Our panel benchmarks widely used representatives of the
matching, qLDPC, and search-based families side by side. Tensor-network
decoders are excluded as impractical at Willow's distances ($d \leq 7$), and
Union-Find as a faster variant of the matching family MWPM already
represents.

Minimum-weight perfect matching (MWPM) is the standard baseline, near-optimal
under graphlike noise models and implemented in
PyMatching~\cite{higgott2022pymatching}, against which most new decoders,
including NVIDIA's~\cite{nvidia_ising2026}, are measured. Around it sit belief
matching~\cite{higgott2023beliefmatching}, belief propagation with
ordered-statistics~\cite{panteleev2021degenerate,roffe2020decoding} or
localized-statistics~\cite{hillmann2024lsd} post-processing for the broader
quantum low-density parity-check (qLDPC) setting, and search-based
decoding~\cite{beni2025tesseract} at the high-accuracy end.
Section~\ref{subsec:panel} defines the panel we assemble from these families
and what each member does with its prior.

\subsection{Synthetic Circuit-Level Noise Models}

Whichever decoder is used, it must be evaluated on a noise model, and in
practice that model is typically synthetic. Circuit-level noise
simulations inject Pauli errors after each gate in a stabilizer circuit
and are commonly generated with Stim~\cite{gidney2021stim}, which also
compiles circuits into the detector error models (DEMs) that decoders
consume. Published models span a range, from generic templates that use no
device data~\cite{gidney_si1000} to models fitted to one target device, either
by estimating per-mechanism circuit parameters~\cite{spitz2018} or by
estimating a DEM from measured syndrome
statistics~\cite{arms2026willowdem,blumekohout2025dem}.
Section~\ref{sec:noise_ladder} defines the four rungs we build across this
range. The paper's central question is whether moving toward the
device-specific end yields more faithful decoder rankings.

\begin{figure*}[!t]
\centering
\includegraphics[width=\linewidth]{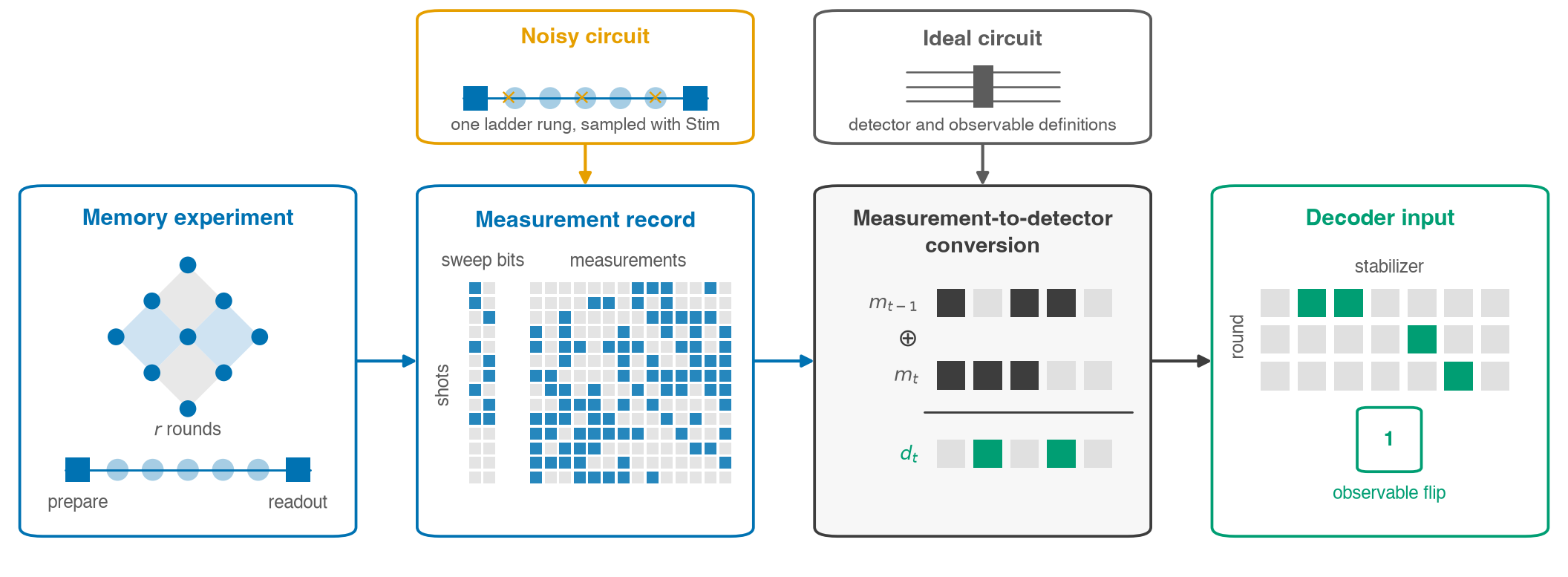}
\caption{Ingest pipeline: a memory experiment on-device is turned into the detection-event lattice a decoder consumes and the observable flip to score against.}
\label{fig:ingest}
\end{figure*}

\subsection{Learned Decoders: AlphaQubit and NVIDIA Ising}

Learned decoders are the newest entrants and the most relevant to this
work. AlphaQubit~\cite{bausch2024alphaqubit}, a recurrent
transformer-based network, was a milestone: trained on a mix of synthetic
and correlation-fit noise models and fine-tuned on real Sycamore data at
distances 3 and 5, it reported lower logical error rates than both MWPM
and a tensor-network decoder on that specific memory experiment. NVIDIA's
Ising pre-decoder~\cite{nvidia_ising2026} pursues the opposite objective:
rather than maximizing accuracy, it uses a convolutional network to reduce
the detection-event lattice before handing off to PyMatching, aiming to
improve the accuracy--latency frontier. The two were validated very
differently, with AlphaQubit against real hardware, but the Ising
pre-decoder in its original release entirely against synthetic
circuit-level noise, with its reported speedups measured only at code
distances ($d \geq 13$) well above those Willow realizes. That difference
in validation is what our benchmark is built to probe. We include
AlphaQubit as context rather than in our panel: it is not open-source in a
reproducible form, and we do not evaluate it on Willow.

\subsection{Calibrating Noise to Real Hardware, and the Sim-to-Real Gap}
\label{subsec:calibration}

If synthetic rankings can mislead, the natural response is to calibrate
the noise model to the device, and a recent line of work does exactly
this, splitting on whether the decoder is in the loop. The decoder-aware
route, taken by Sivak, Newman, and Klimov~\cite{sivak2024decoderpriors},
uses a reinforcement-learning-inspired method to tune decoder priors so as
to minimize an MWPM decoder's logical error rate, on repetition and
surface code experiments on Sycamore; the ``MWPM-RL'' entry in our own panel decodes with MWPM from a DEM
produced by this method~\cite{sivak2024decoderpriors}, rather than from a
generic prior. The decoder-free route, taken by Arms et
al.~\cite{arms2026willowdem} and by Blume-Kohout and
Young~\cite{blumekohout2025dem}, estimates DEMs directly from syndrome
statistics with no decoder involved. On the 72- and 105-qubit Willow chips,
Arms et al. further identify structure a DEM cannot capture at all,
including correlated flips of adjacent detectors across many rounds and
signatures consistent with radiation events. Together these establish two
points our work builds on: that device-calibrated noise can differ
materially from generic synthetic noise, and that the decoder-aware and
decoder-free calibrations need not agree with each other. What they leave
open, and what we address, is the downstream consequence for \emph{decoder
ranking}: whether these differences change which decoder a practitioner
would choose, and how that answer shifts with code distance and round
count.

\section{Methods}
\label{sec:methods}

We run every decoder on the same circuits five times, once on each of the four
synthetic noise models and once on Willow itself, scored each time by the
logical error rate.

\subsection{Ingest Pipeline}
The device supplies only the measurement record and the sweep bits. Everything
else in the dataset, detection events and observable flips included, is
recomputed from them rather than read as shipped. The pipeline in
Fig.~\ref{fig:ingest} produces a decoder's input, the detection events, and its
answer key, the observable flip.

Detection events come from the ideal circuit, a noiseless circuit in which each
detector fixes the parity of a set of measurement outcomes when the circuit runs
without error. Measurement-to-detector conversion evaluates those parities on
the device record, and a detector takes the value $1$ wherever the record breaks
its parity.

A detector combines different outcomes depending on where it sits in the
circuit: the opening layer compares a stabilizer against the value the prepared
state fixes, a bulk detector compares a stabilizer against its own value one
round earlier, and the closing layer compares the final stabilizer measurement against
the matching parity of the data-qubit readout. The ideal circuit defines the
detectors and contains no measurement outcomes.

Detection events are well defined only when the sweep bits are read alongside
the measurement record. Each sweep bit applies a conditional $X$ to one data
qubit at preparation, randomizing the initial state per shot. Randomizing the
input moves the fixed value of every detector it affects, and the fixed value
of the logical observable with it. Read without the sweep bits, the same record
yields detection events and truth bits for a state the device never prepared.

At distance $d$ over $r$ rounds the circuit defines $(d^2-1)/2$ stabilizers of
each type and $r(d^2-1)$ detectors, so a shot yields that many bits plus one
observable bit. That observable bit is the only ground truth hardware offers,
and it is what every decoder is scored against: the physical error pattern
behind it is unrecoverable.

Synthetic data enters the same pipeline one stage earlier, as a second source
for the measurement record in Fig.~\ref{fig:ingest}. Each rung of
Section~\ref{sec:noise_ladder} samples a noisy circuit, and its records pass
through the same conversion and the same scoring, with Stim supplying the
injected logical flip as ground truth. The logical error rates are comparable
because real and synthetic runs differ in only the origin of the measurement
record.

\subsection{The Willow Below-Threshold Dataset}
All real data comes from below-threshold memory experiments on the Willow
105-qubit superconducting processor~\cite{willow2024}. Each configuration prepares a logical qubit in a
known $X$ or $Z$ eigenstate, holds it for $r$ rounds of rotated-surface-code
stabilizer measurement, and reads it out, repeated for 50{,}000 shots. The
dataset covers three distances, $d=3$, $5$, and $7$, on 14 physical patches: nine
at $d3$, four at $d5$, and one at $d7$. With two bases and 15 round counts the
corpus totals 420 configurations. The patch count falls steeply with distance,
which limits how well a device-calibrated noise model can be conditioned at
$d7$, where a single patch supplies every statistic.

\subsection{The Noise-Fidelity Ladder}
\label{sec:noise_ladder}
The noise-fidelity ladder is an ablation over noise models. Four synthetic
models of increasing fidelity to the device generate detection events for the
same circuits that produced the real data, and each rung is then decoded by the
same panel and scored by the same rule as the real data. The noise model is the
only thing that differs between rungs.

The four rungs, ordered by the amount of device information they use, are:

\begin{enumerate}
\item \textbf{Uniform depolarizing.} A single error rate applies to every
operation in the circuit, so the model has no way to express the
order-of-magnitude spread across operation types on Willow, where two-qubit
gates and measurement fail far more often than idling~\cite{willow2024}. The
circuit is unchanged, so every edge of the detector graph is still correct and
only the weights are wrong, each off by the factor separating that operation's
true error rate from the global average.
\item \textbf{SI1000.} The standard circuit-level noise model for
superconducting hardware~\cite{gidney_si1000}, also Willow's reference model,
with per-operation error rates in fixed ratios characteristic of real devices
and scaled by one parameter $p$.
\item \textbf{25-parameter fit.} A circuit-level model with 25 physically
grouped parameters (2 state-preparation, 2 measurement, 3 idle-during-gate,
3 idle-during-SPAM, and 15 two-qubit-gate Pauli rates), fitted to each patch
from its detection-event fractions and its pairwise detector-correlation
($p_{ij}$) matrix~\cite{spitz2018}. Two of the 25 are not independent on this
device. Willow prepares and measures in the $X$ basis by conjugating the
$Z$-basis operations with Hadamard gates, so both bases run through the same
physical preparation and the same physical measurement, and no syndrome
statistic can separate an $X$-basis rate from its $Z$-basis counterpart.
The model has 23 free parameters on this compilation, because
$p_{\mathrm{prep},X}=p_{\mathrm{prep},Z}$ and
$p_{\mathrm{meas},X}=p_{\mathrm{meas},Z}$. The model cannot express any $X$/$Z$
asymmetry in preparation or measurement, and Section~\ref{sec:xzasym} measures
that asymmetry on Willow.
\item \textbf{Syndrome-estimated DEM.} A detector error model recovered edge by
edge for each configuration by inverting the same $p_{ij}$ statistics,
following the syndrome-DEM
estimators of~\cite{arms2026willowdem,blumekohout2025dem}. This DEM can take
edge weights that no setting of the 25 parameters produces, because each edge
probability is estimated independently.
\end{enumerate}

The four rungs are split into two types, uncalibrated and calibrated. Uniform
and SI1000 are uncalibrated: their rates are constants of the model definition,
a single global $p$ and fixed multiples of $p$, chosen without any Willow
measurement. The 25-parameter fit and the syndrome-estimated DEM are calibrated:
their rates are inferred from Willow shots, so they alone are subject to
sampling error and ill-conditioning.

The estimators read detection-event frequencies and detector correlations only.
Logical outcomes, from which every logical error rate and every ranking in this
paper is computed, never enter a fit. Each calibrated rung is then evaluated on
fresh samples drawn from its own noise model, not on the device shots used to
estimate it. Neither rung is therefore scored on data that shaped it.

\subsection{The Decoder Panel}
\label{subsec:panel}

The panel contains six decoders, each reading a detector error model as its
prior: for a synthetic rung its own noise model's, for Willow the reference
SI1000 one.

\begin{enumerate}
\item \textbf{PyMatching (MWPM).} Minimum-weight perfect matching over the
matching graph~\cite{higgott2022pymatching}, with every error probability fixed
as an edge weight before the decode begins.
\item \textbf{BeliefMatching.} Belief propagation first, then matching on a
graph reweighted by the posteriors it
produces~\cite{higgott2023beliefmatching}.
\item \textbf{BP+OSD.} Belief propagation from the \texttt{ldpc} library, with
the remaining ambiguity resolved by ordered-statistics
post-processing~\cite{panteleev2021degenerate,roffe2020decoding}.
\item \textbf{BP+LSD.} The same, with localized-statistics
post-processing~\cite{hillmann2024lsd}.
\item \textbf{Tesseract.} An $A^\ast$ search over candidate error sets, run on
the full hyperedge structure with no reduction to a
graph~\cite{beni2025tesseract}.
\item \textbf{Ising pre-decoder.} NVIDIA's network in front of PyMatching
(Section~\ref{subsec:adapter}). It is the one entry that does not read the
prior, which enters only at the matching step.
\end{enumerate}

Two of the six run in more than one configuration, giving eight: the Ising
pre-decoder appears once per released model size, and MWPM appears again as
MWPM-RL, reading the RL-optimized DEM of Section~\ref{subsec:calibration}. No
synthetic noise model produces such a prior, so MWPM-RL is evaluated only on
Willow and is excluded from the rank comparisons.

\subsection{The Ising Input Mapping}
\label{subsec:adapter}

Ising hands PyMatching a residual syndrome, and the reported flip is the
modulo-two sum of the network's own frame with PyMatching's decode of that
residual~\cite{nvidia_ising2026}.

Running this model on Willow data requires an adapter. Willow detection events
are stored as a flat $(\text{shots}, \#\text{detectors})$ array in the emission
order of the device's XZZX-variant circuit, and the model expects a
$(4, T, D, D)$ tensor ordered as in the CSS-style synthetic training
circuits. The four channels hold the two syndrome types and their
stabilizer-presence masks, $T$ is the number of rounds, and $D$ is the code
distance.
Our adapter reorders detectors by a
$(\text{layer}, \text{stabilizer type}, x, y)$ key shared with a reference
synthetic circuit and then applies the released
\texttt{dets\_to\_predecoder\_inputs} transform.

Four lattice orientations place Willow's XZZX stabilizers on the CSS grid.
Each is named by the stabilizer type, X or Z, of the grid's first bulk
syndrome qubit and by the orientation, vertical (V) or horizontal (H), of its
northwestern boundary, which fixes the direction of the logical operator: XV,
XH, ZV, and ZH. Nothing in the detection events identifies which one the model
expects. We use XV throughout and test that choice against the other
three in Appendix~\ref{app:isingvalid}. The two released models have
receptive fields $R=9$ and $R=13$, so every Ising result is reported with its $d$ and $R$.

\subsection{Scoring}
\label{subsec:scoring}

A decoder receives the syndrome and predicts the logical flip, and the logical
error rate is the fraction of shots whose prediction differs from the recorded
flip. Cumulative rates are converted to per-cycle rates through the fidelity
inversion
$\varepsilon_{\mathrm{cycle}} = \tfrac12\bigl(1-(1-2\varepsilon)^{1/r}\bigr)$,
where $\varepsilon$ is the cumulative rate over $r$ rounds, the convention of
the dataset release~\cite{willow2024}.

\subsection{Metrics}
\label{subsec:metrics}

Each configuration is stored as a logical-error count and a bit-packed vector
of per-shot outcomes. The rate follows from the count, and its confidence
interval from resampling that count under the binomial. Rank-flip
probabilities come from the joint bootstrap of Appendix~\ref{app:expdetails}.

Rank agreement between a rung and Willow is quantified per distance by
Kendall's $\tau$~\cite{kendall1938} with its $p$-value, computed over the
decoders present in both sources (Table~\ref{tab:tau_ladder}). $\tau = 1$ indicates
identical orderings and $\tau = 0$ indicates agreement no better than a random
permutation. All $p$-values
are reported.

\section{Experiments}
\label{sec:experiments}

An evaluation runs one decoder on one configuration of one data source, a
distance, basis, patch, and round count. Table~\ref{tab:run_matrix} counts
all 5{,}934 of them. MWPM runs everywhere. We cap the costlier decoders at
$r = 70$ on Willow and $r = 30$ on the synthetic rungs, with fewer shots.
Fewer shots cause wider confidence intervals. Appendix~\ref{app:expdetails}
lists the ten cells that deviate from the table.

\begin{table}[H]
\centering
\caption{Evaluations per decoder. Each cell is configurations $\times$ shots
per configuration, on Willow and on each of the four synthetic rungs.}
\label{tab:run_matrix}
\begin{tabular*}{\columnwidth}{@{\extracolsep{\fill}}lcc}
\toprule
 & Willow & Synthetic \\
\midrule
MWPM & 420 $\times$ 50k & 420 $\times$ 20k \\
MWPM-RL & 420 $\times$ 50k & --- \\
BP+LSD & 168 $\times$ 50k & 112 $\times$ 20k \\
Ising-fast & 140 $\times$ 50k & 84 $\times$ 20k \\
Ising-accurate & 140 $\times$ 50k & 84 $\times$ 20k \\
BeliefMatching & 168 $\times$ 10k & 112 $\times$ 5k \\
BP+OSD & 168 $\times$ 10k & 112 $\times$ 5k \\
Tesseract & 168 $\times$ 10k & 112 $\times$ 5k \\
\bottomrule
\end{tabular*}
\end{table}

We run every decoder on rounds $2 \le r \le 30$ of every source, so
cross-source comparisons average per-cycle rates over that range. The
$r \le 30$ cutoff keeps cumulative rates far below $0.5$, so the inversion of
Section~\ref{subsec:scoring} stays well conditioned.

\section{Results}

\subsection{Rank Agreement}
\label{sec:rank_ladder}

Each rung yields a decoder ranking. We order the seven panel entries by mean
per-cycle logical error rate over rounds 2--30 at fixed code distance, and
compute the same ordering from the real Willow data. MWPM-RL decodes from a device-tuned prior with no synthetic
counterpart and is excluded from cross-comparisons.

\begin{figure}[!h]
\centering
\includegraphics[width=\columnwidth]{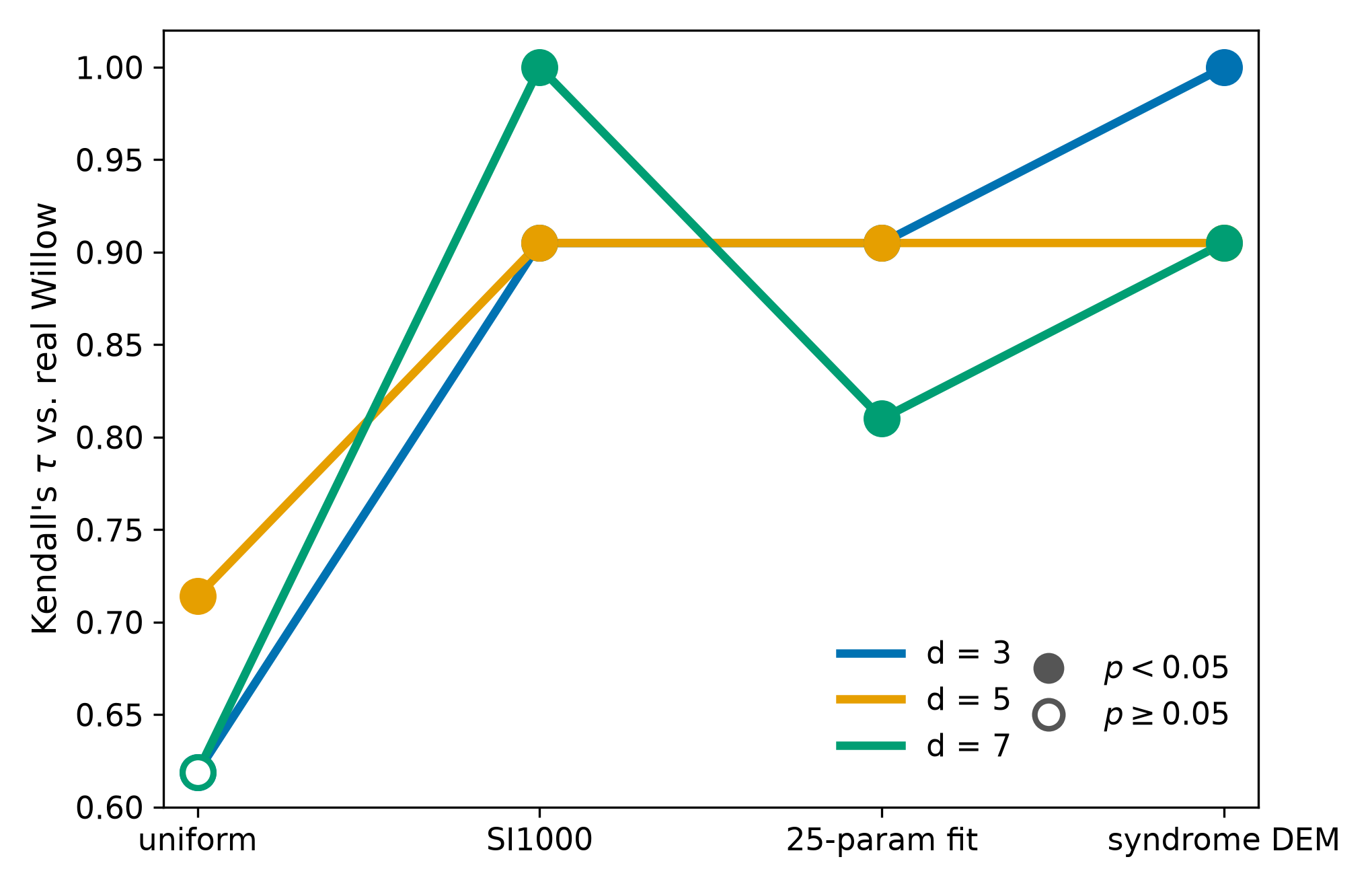}
\caption{Kendall's $\tau$ between each synthetic noise model's decoder ranking
and the real-Willow ranking, by code distance, over the rounds 2--30 window.
$\tau = 1$ denotes an identical ordering. Filled markers indicate agreement
significant at $p < 0.05$ against a random ordering; open markers indicate
agreement that does not reach significance, both under uniform depolarizing
noise.}
\label{fig:tau_ladder}
\end{figure}

Figure~\ref{fig:tau_ladder} and Table~\ref{tab:tau_ladder} give these values.
Uniform noise agrees weakly with real hardware: $\tau = 0.619$,
$0.714$, and $0.619$ at $d = 3$, $5$, and $7$, and neither the $d = 3$ nor the
$d = 7$ value reaches significance ($p = 0.069$ for both). The three rungs carrying circuit-level
structure clear $\tau \geq 0.810$ in all nine rung--distance cells and reach
exact agreement in two. Uniform noise is separated from the other three rungs
by a wider margin than separates those three from each other.

At $d = 7$ the generic
SI1000 template reaches $\tau = 1.000$, while the 25-parameter fitted model
reaches $0.810$ and the syndrome-estimated DEM reaches $0.905$. Neither
$d = 3$ nor $d = 5$ shows this inversion.
With seven panel entries the comparison spans
$\binom{7}{2} = 21$ pairs, which bounds the power of the significance test.
The uniform rung failing $p < 0.05$ where the structured rungs clear it
carries more weight than any single $p$-value.

\subsection{Rank Movement}

\begin{figure*}[!h]
\centering
\includegraphics[width=\linewidth]{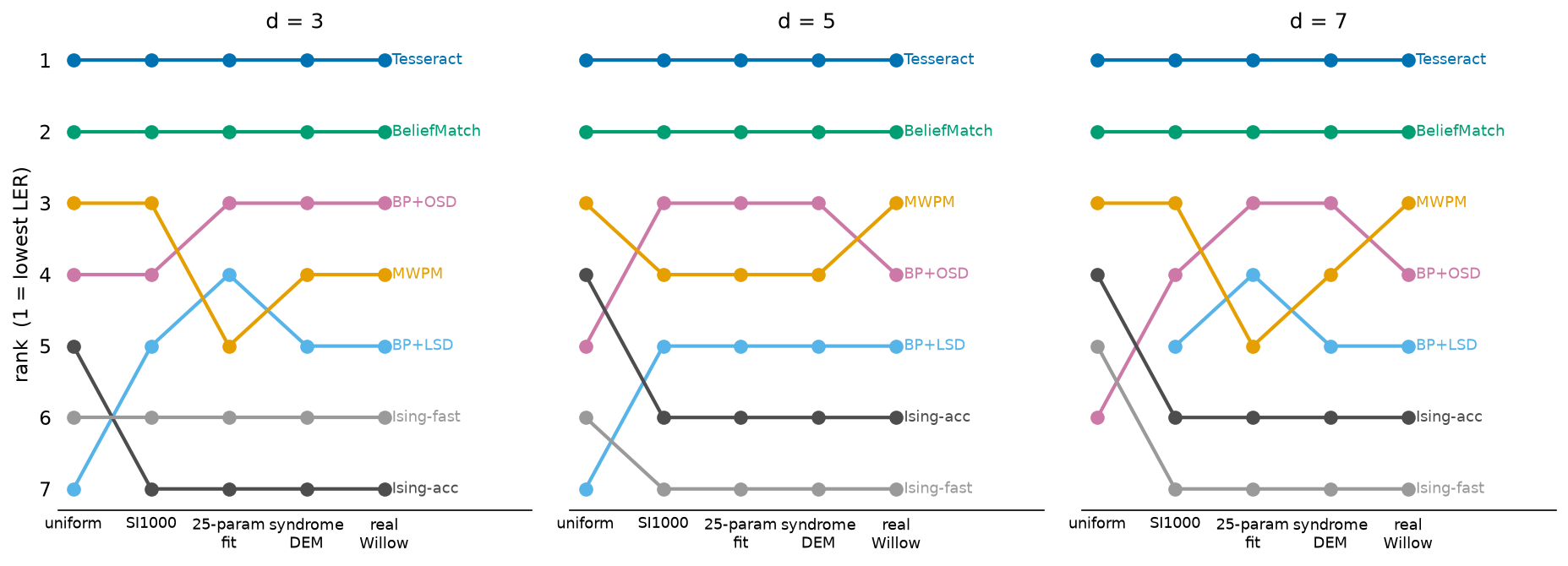}
\caption{Decoder rank under each noise model and on real Willow, by code
distance. Rank 1 is the lowest per-cycle logical error rate.}
\label{fig:rank_bump}
\end{figure*}

The $\tau$ values in Section~\ref{sec:rank_ladder} measure how much a ranking
disagrees with real hardware but not which decoders cause the disagreement. We
therefore track the rank each decoder takes under every noise model and on real
Willow, at each code distance as shown in Figure~\ref{fig:rank_bump}.

\label{sec:rank_movement}

Tesseract ranks first and BeliefMatching second in all fifteen
model--distance cells, and the two Ising variants rank sixth and seventh on the
three structured models and on Willow. Under uniform noise BP+LSD drops below
both Ising variants at every distance, and BP+OSD below at least one at
$d \geq 5$. All other movement happens among MWPM, BP+OSD, and BP+LSD. MWPM and BP+OSD differ by
$6.6\times10^{-4}$ per cycle on real Willow
at $d = 7$, across a panel that spans a factor of 16. The noise model decides
the ranking only where the decoders are already near-tied.

\begin{figure*}[!h]
\centering
\includegraphics[width=\linewidth]{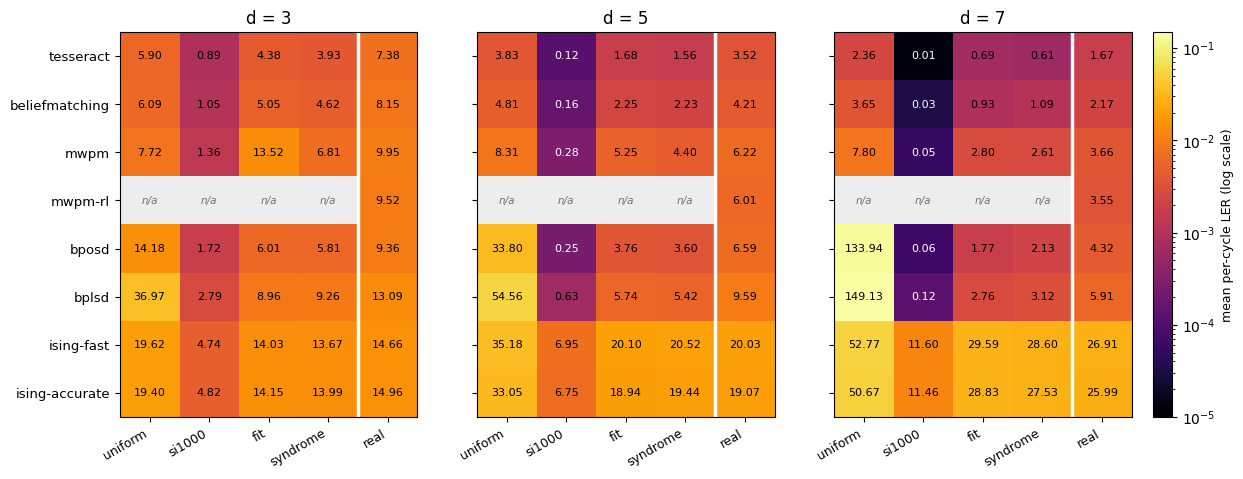}
\caption{Mean per-cycle logical error rate by decoder, noise model, and code
distance, over the rounds 2--30 window.}
\label{fig:ler_ratio}
\end{figure*}

\subsection{Rank Stability}
\label{sec:rank_stability}

The fitted rung's 23 parameters are estimates, each with its own error. To
test how much of its ranking survives that error, we double one parameter,
regenerate the rung, and rank the panel again, once for every parameter
(Appendix~\ref{app:expdetails}).

The ranking barely moves. Ten of the 23 parameters reorder nothing, and the
rest reorder at most one adjacent pair at any distance. Eight of the nine
parameters carrying a Pauli $Y$ reorder the panel, and only five of the
remaining fourteen do. The three that reorder it at every distance are
\texttt{p\_idle\_spam\_Y}, \texttt{p\_idle\_cnot\_Y}, and
\texttt{p\_cnot\_ZY}. Every swap is MWPM with BP+LSD, BP+OSD with MWPM, or the
two Ising variants.

Parameter error is systematic, sampling error is not, so we re-rank the panel
on 40 refits of resampled device shots as well
(Appendix~\ref{app:expdetails}). Tesseract ranks first in every draw at every
distance. Reordering below it grows sharply with code distance
(Table~\ref{tab:tab_bootstrap_flips}). The same three pairs reorder here.

\begin{table}[t]
\centering
\caption{Fraction of the 40 bootstrap refits in which a decoder pair swaps
rank, listed by pair with their union in the last row.}
\label{tab:tab_bootstrap_flips}
\begin{tabular}{lrrr}
\toprule
Pair & $d=3$ & $d=5$ & $d=7$ \\
\midrule
MWPM $\leftrightarrow$ BP+LSD & 0.00 & 0.20 & 0.38 \\
BP+OSD $\leftrightarrow$ MWPM & 0.00 & 0.00 & 0.25 \\
Ising-fast $\leftrightarrow$ Ising-acc. & 0.03 & 0.00 & 0.00 \\
\addlinespace
Any pair & 0.03 & 0.20 & 0.62 \\
\bottomrule
\end{tabular}
\end{table}

\subsection{Absolute Error}
\label{sec:absolute_ler}

Rank agreement is insensitive to the scale of a model's error rates. We
therefore report per-cycle logical error rate directly.

Figure~\ref{fig:ler_ratio} shows that the two calibrated models sit closest to
hardware, while SI1000 underestimates by one to more than two orders of magnitude and
worsens with distance, reaching $1\times10^{-5}$ for Tesseract at $d = 7$
against $1.67\times10^{-3}$ on hardware. Uniform noise stays near hardware for
most decoders but diverges sharply for BP+OSD and BP+LSD, reaching
$1.34\times10^{-1}$ and $1.49\times10^{-1}$ at $d = 7$. MWPM-RL is absent from
the synthetic models, having no synthetic counterpart.

\subsection{Accuracy and Latency}
\label{sec:pareto}

Decoder selection depends on decode time as well as accuracy. We place each
decoder at its median per-shot latency of a batched decode and its median
per-cycle logical error rate on real Willow. A decoder reaches the Pareto
frontier when no other decoder is both faster and more accurate. Latency is
measured over the same rounds 2--30 window as the error rates, so every decoder
is timed on the same circuit depths (Table~\ref{tab:tab_latency}).

\begin{figure*}[!h]
\centering
\includegraphics[width=\linewidth]{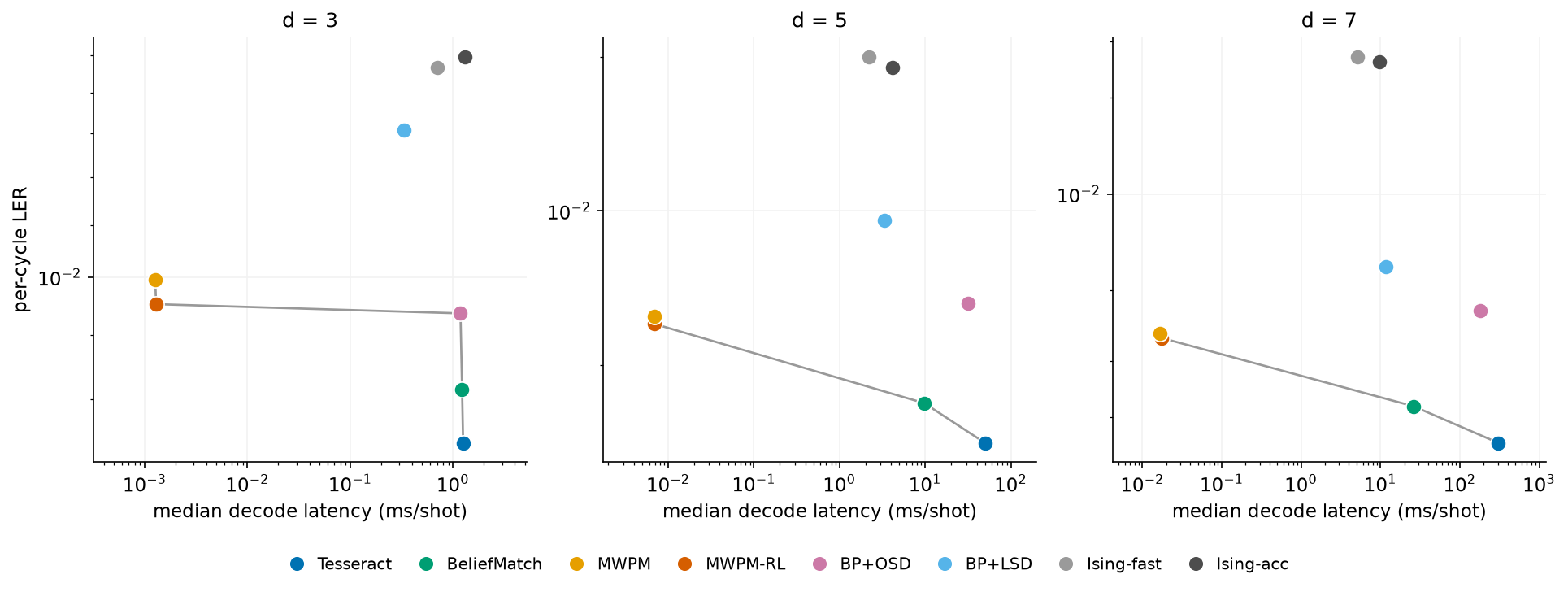}
\caption{Per-cycle logical error rate against decode latency on  Willow, by
code distance. The line marks the Pareto frontier.}
\label{fig:pareto}
\end{figure*}

\begin{table}[t]
\centering
\caption{Median decode latency (ms/shot) on real Willow per decoder $\times$ distance, with fitted growth per +2 code distance and the extrapolated d=9 cost.}
\label{tab:tab_latency}
\begin{tabular}{lrrrrr}
\toprule
Decoder & d=3 & d=5 & d=7 & growth /+2d & extrap. d9 \\
\midrule
Tesseract & 1.253 & 49.842 & 305.501 & 15.6x & ~4.8 s \\
BeliefMatch & 1.225 & 9.748 & 26.052 & 4.6x & ~120 ms \\
MWPM & 0.001 & 0.007 & 0.017 & 3.6x & ~0.061 ms \\
MWPM-RL & 0.001 & 0.007 & 0.017 & 3.7x & ~0.064 ms \\
BP+OSD & 1.181 & 31.478 & 182.085 & 12.4x & ~2.3 s \\
BP+LSD & 0.336 & 3.335 & 11.744 & 5.9x & ~69 ms \\
Ising-fast & 0.705 & 2.220 & 5.086 & 2.7x & ~14 ms \\
Ising-acc & 1.311 & 4.167 & 9.609 & 2.7x & ~26 ms \\
\bottomrule
\end{tabular}
\end{table}

MWPM and MWPM-RL sit together at the low-latency end of the frontier at every
distance, followed by BeliefMatching and then Tesseract
(Fig.~\ref{fig:pareto}). The two matching entries are separated by less than a
microsecond per shot, so their order along the frontier is arbitrary. BP+OSD
reaches the frontier at $d = 3$ alone. MWPM-RL is both faster and more accurate
than BP+LSD at all three distances.

\subsection{Ising Analysis}
\label{sec:ising}

Another panel decoder matches or improves on the Ising pre-decoder in both
per-cycle error rate and decode latency in 278 of the 280 Willow evaluations.
The two remaining evaluations are the same configuration, patch
\texttt{q8\_5} in the $Z$ basis at $d = 3$ and $r = 30$, once for each model
size. MWPM decodes faster at every distance: at $d = 7$ it takes $0.017$~ms per
shot against $5.09$~ms for Ising-fast and $9.61$~ms for Ising-accurate
(Table~\ref{tab:tab_latency}), and reaches $3.66\times10^{-3}$ per cycle
against $2.69\times10^{-2}$ and $2.60\times10^{-2}$. Ising-accurate's error
rate stays within a few percent of Ising-fast's at every distance, so the
paragraphs below quote Ising-fast.

Ising is the only decoder in the panel whose error rate rises with code
distance on real Willow, and the only one that rises on every source. Its
per-cycle rate on hardware is $1.47\times10^{-2}$ at $d = 3$ and
$2.69\times10^{-2}$ at $d = 7$, a factor of $1.84$, while every other decoder
improves over the same span and MWPM falls by a factor of $2.72$. The rise repeats on all four synthetic models, from $2.09$
on the syndrome-estimated DEM to $2.69$ under uniform noise. Both released
models were trained for receptive fields wider than the codes measured here,
$R = 9$ and $R = 13$ against $d \leq 7$.

\begin{figure*}[!h]
\centering
\includegraphics[width=\linewidth]{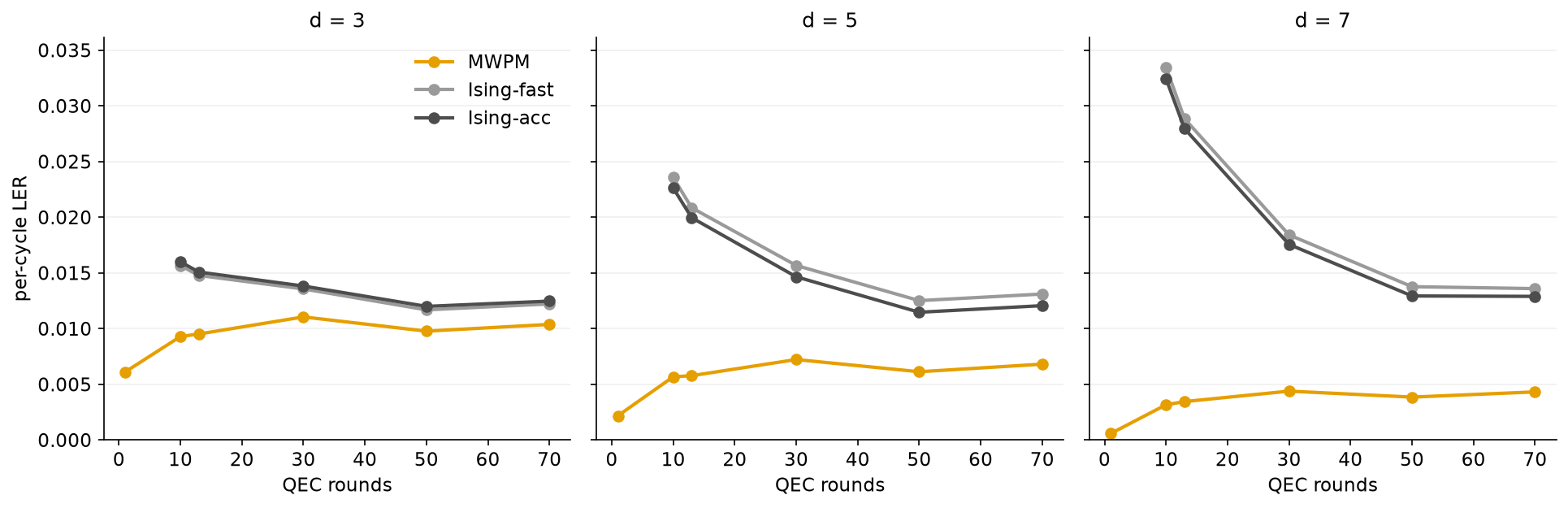}
\caption{Per-cycle logical error rate against round count on Willow, by
code distance.}
\label{fig:ising_rounds}
\end{figure*}

Ising's error rate relative to MWPM depends on circuit depth. At $d = 7$ the
Ising-fast per-cycle rate is $10.6$ times MWPM's at $r = 10$ and $3.2$ times at
$r = 70$, while MWPM stays between $3.2\times10^{-3}$ and $4.4\times10^{-3}$
across that span (Fig.~\ref{fig:ising_rounds}). A per-cycle rate that falls as
rounds increase follows from a fixed per-shot error contribution divided over
more cycles, so most of Ising's error is already present at the shallowest
depth measured. Any single multiple quoted against MWPM therefore depends on
the round count, and the rounds 2--30 window used throughout this paper gives
$7.35$ at $d = 7$.

All four synthetic models reproduce this depth dependence, and the two
calibrated models reproduce it most strongly. At $d = 7$ and $r = 10$ the
Ising-fast per-cycle rate is $3.35\times10^{-2}$ on hardware,
$3.75\times10^{-2}$ on the fitted model and $3.50\times10^{-2}$ on the
syndrome-estimated DEM. Of the five sources, only SI1000 produces a lower Ising
error rate at that setting.

Ising's error rate responds little to the noise model. Across the five sources
its per-cycle rate spans a factor of $4.1$, $5.1$ and $4.5$ at $d = 3$, $5$ and
$7$, against $9.9$, $30.1$ and $154.0$ for MWPM and $8.2$, $136.4$ and $2107$
for BP+OSD. Ising is the one panel entry that does not read the prior
(Section~\ref{subsec:panel}), and the noise model reaches it only through the
matching step applied to the residual.


\section{Discussion}

\subsection{Why Calibration Does Not Guarantee Rank Agreement}
\label{sec:disc_conditioning}

Section~\ref{sec:rank_ladder} showed the two calibrated models losing the rank
comparison at $d = 7$ to a generic template, while
Section~\ref{sec:absolute_ler} showed the same models coming closest to
hardware in absolute error. Both are estimated from device data, so each edge
parameter is only as reliable as the data supporting it.

\begin{figure}[t]
\centering
\includegraphics[width=\columnwidth]{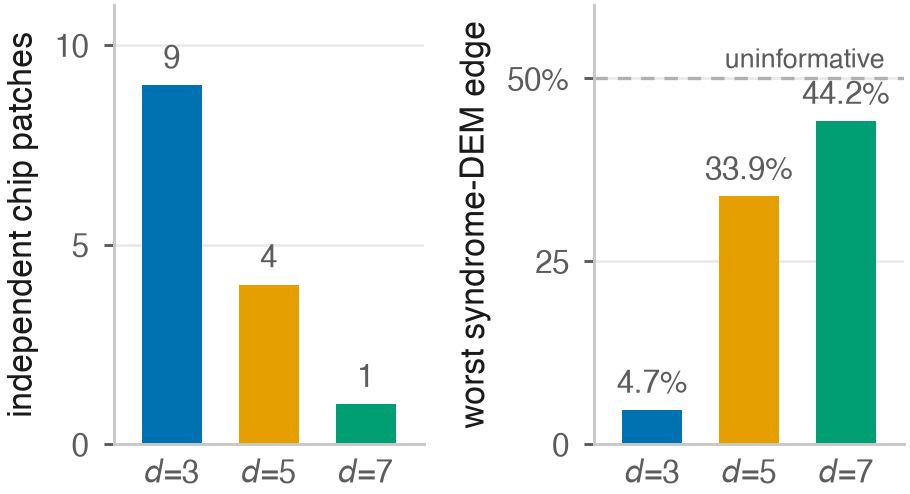}
\caption{Independent code patches per distance in the Willow corpus (left) and
the largest edge probability in the syndrome-estimated DEM at each distance (right). An edge
probability of $0.5$ carries no information about whether that error occurred.}
\label{fig:conditioning}
\end{figure}

Figure~\ref{fig:conditioning} shows the corpus
holds nine independent patches at $d = 3$, four at $d = 5$, and one at
$d = 7$, while the decoding graph grows with $d^{2}$. Each estimated parameter
receives less data as distance increases, which can be seen in the syndrome-estimated DEM, whose largest edge
probability  reaches $0.442$ at $d = 7$.

MWPM assigns each edge a cost of $-\log\left(p/(1-p)\right)$ and selects the
minimum-weight matching consistent with the
syndrome~\cite{higgott2022pymatching}. An edge estimated near $0.5$ therefore
carries a cost near zero, and paths through it become nearly free irrespective
of the syndrome. The 25-parameter fit, whose rates are all below $10^{-2}$,
cannot produce such an edge, and its two swapped pairs at $d = 7$ have a
different origin. It places BP+OSD ahead of MWPM at every distance, by ratios
of 0.44, 0.72, and 0.63 against 0.94, 1.06, and 1.18 on Willow, and its MWPM
and BP+LSD rates at $d = 7$, $2.80\times10^{-3}$ and $2.76\times10^{-3}$, are
a near-tie whose flip probability Table~\ref{tab:tab_bootstrap_flips} puts at
0.38. SI1000~\cite{gidney_si1000} contains
no estimated parameters, is unaffected by patch count, and preserves the
hardware ordering where the calibrated models lose it.

Thin data is not the only limit.
\label{sec:xzasym}
Willow prepares and measures the $X$ basis by conjugating the $Z$ operations,
so both bases share one preparation rate and one measurement rate in the
fitted model (Section~\ref{sec:noise_ladder}). No quantity of shots separates
them.

The device does distinguish them. On Willow the $X$ basis is worse in all 24
decoder--distance cells, by 4\% to 39\%, and configuration by configuration
$Z$ is the lower of the two in most matched pairs for every decoder, from 29
of 42 for Ising-accurate to 39 of 42 for BP+LSD. The synthetic rungs are
symmetric, with ratios scattering above and below one, and on the fit rung the
count reverses for BeliefMatching and Tesseract, to 12 and 11 of 42
(Table~\ref{tab:tab_xz_asymmetry}). A model that sits closest to hardware in
absolute error can still miss a difference the device shows at every distance.

\begin{table}[t]
\centering
\caption{X$>$Z basis asymmetry (mean per-cycle LER ratio, MWPM). Real Willow is X$>$Z in 24/24 cells (mean +13.6\%, ratio 1.04-1.39); the fit rung reproduces it in only 14/21 cells (median ratio ~1.02) and syndrome in 11/21 -- neither Pauli rung reproduces the device asymmetry.}
\label{tab:tab_xz_asymmetry}
\begin{tabular}{lrrr}
\toprule
Distance & real X/Z & fit X/Z & syndrome X/Z \\
\midrule
d=3 & 1.107 & 1.055 & 0.979 \\
d=5 & 1.124 & 0.981 & 0.934 \\
d=7 & 1.162 & 0.947 & 0.981 \\
\bottomrule
\end{tabular}
\end{table}

Existing work on device noise estimation targets agreement with the device.
Sivak et al.~\cite{sivak2024decoderpriors} tuned decoder priors to minimize
MWPM's logical error rate; Arms et al.~\cite{arms2026willowdem} and
Blume-Kohout and Young~\cite{blumekohout2025dem} estimated detector error
models from syndrome statistics without a decoder in the loop. Our results agree with that aim and expose a second objective, which is that a model can sit close to the device in absolute error yet still
misrank decoders. Arms et al.~\cite{arms2026willowdem} report a related split,
finding that DEMs estimated from syndromes match unseen syndromes better than
DEMs tuned for logical performance, while the tuned models serve better as
decoder priors.

An error mechanism specific to the single $d = 7$ patch would produce the same
inversion, and this corpus cannot rule that out. The largest edge probability
increases at every distance, though, not only at $d = 7$.

\subsection{Prior Sensitivity as a Selection Criterion}
\label{sec:disc_prior}

Under uniform noise at $d = 7$, BP+OSD and BP+LSD reach $31.0\times$ and
$25.2\times$ their real-hardware error, while MWPM, Tesseract and
BeliefMatching stay within $2\times$. Belief propagation passes messages
weighted by the prior~\cite{panteleev2021degenerate,roffe2020decoding}; a flat
prior leaves those messages unable to distinguish one fault mechanism from another. The stage that follows belief
propagation~\cite{hillmann2024lsd} can only choose among the candidates it receives, so it inherits the prior. MWPM is less exposed because its answer depends
on distances in the matching graph which a flat prior leaves unchanged.
Failures therefore have two causes. A globally flat prior removes the information belief propagation
depends on. A few degenerate edges in an otherwise accurate model leave the
prior informative but corrupt specific matchings, as
Section~\ref{sec:disc_conditioning} described. Each decoder is sensitive to a
different path from a correct prior. 
The sweep of Section~\ref{sec:rank_stability} measures this difference
directly. Doubling \texttt{p\_idle\_spam\_Y} raises MWPM's error
$1.57\times$ against BP+LSD's $1.23\times$, closing a gap the two then cross.
Doubling \texttt{p\_cnot\_ZZ} moves them the other way, $1.43\times$ against
$1.53\times$, and the gap widens. A reordering follows from unequal response
to a parameter, not from the size of the error it adds.

Decoder comparisons typically report accuracy under a matched prior, which
cannot distinguish a decoder that reaches lower error under a good prior and
loses an order of magnitude under a poor one from a decoder that does neither.

\subsection{Interpreting the Ising Result}
\label{sec:disc_ising}
Section~\ref{sec:ising} placed the Ising pre-decoder last on real Willow, and
Section~\ref{sec:rank_movement} placed it last under the three structured
synthetic models as well. Only uniform noise moves it, and only because BP+LSD
and, at larger distance, BP+OSD fall behind it there. Here, the structured
benchmarks got the ranking right because the network reads no prior
(Section~\ref{subsec:panel}), so the noise model barely changes its error rate
and its rank moves only when another decoder's does. Rank agreement obtained this way does not measure the
benchmark's fidelity. It says the decoder is insensitive to what varies in the
ladder.

How much of the hardware result belongs to hardware is a separate question. If
leakage, crosstalk or readout behavior were responsible, the degradation would
appear only on the device, since none of those effects exists in a Pauli
simulation. Section~\ref{sec:ising} finds it on all four synthetic models
instead, and finds the two calibrated models producing a higher Ising error
rate than the device itself. A pre-decoder trained on one lattice geometry and
evaluated through a mapping onto another would incur a penalty that is
independent of the noise model and grows with the lattice, which is what these
measurements show.

The evaluation conditions are part of what a learned decoder's reported margin
means. AlphaQubit~\cite{bausch2024alphaqubit} was fine-tuned on the device it
was tested on and run in its native geometry, and its margin over MWPM was
measured under both. Ising was trained on synthetic noise for a different
layout, and our evaluation preserves neither.

\subsection{Limitations}
\label{sec:disc_limits}

The corpus offers one patch at $d = 7$, so distance and available data vary
together, and we cannot tell whether poor conditioning or code distance itself
drives the inversion. Separating them requires holding one fixed while varying
the other, which we leave as a future experiment.

Seven panel entries give 21 rank pairs, so $\tau$ moves in steps of roughly $0.095$
and every discordant pair is a visible jump. Detecting agreement is therefore
easy, while comparing degrees of it is not, since the gap between
$\tau = 0.905$ and $\tau = 1.000$ amounts to one swapped pair among two closely
spaced decoders. What the $p$-values distinguish is agreement from chance
rather than one noise model from another. The SI1000 cell at $d = 7$ also
rests on few errors: Tesseract, BeliefMatching, and BP+OSD record 6, 17, and
32 logical errors over their rounds 2--30 configurations, so its
$\tau = 1.000$ is the least certain value in Table~\ref{tab:tau_ladder}.

Four noise models do not exhaust the space. All four are stochastic Pauli
models, so none of them represents the correlated structure Arms et
al.~\cite{arms2026willowdem} report on this device, and a model built on
different principles could rank differently.

\subsection{Practical Implications}
\label{sec:disc_practice}

Selecting a decoder depends on whether the noise model has circuit-level
structure, not on how closely it matches the device. Flat depolarizing noise
misranks, while every model with per-operation structure recovers the
hardware ordering to within two swapped pairs, including the generic template
built from no device data. Fitting to the device improves absolute error rates and returns nothing in rank
agreement. At $d = 7$, where one patch supplies every statistic, both fitted
models agree with the hardware ordering less well than the generic template.

Therefore, estimation methods that report per-edge uncertainty
rather than a point estimate would let a decoder discount an unreliable edge,
which tests directly whether ill-conditioning causes the $d = 7$ inversion.
Repeating the comparison to a device with several patches at high distance would
show whether the failure we attribute to data scarcity disappears when the data
is there.

\section{Conclusion}

This work established, on below-threshold hardware, that a synthetic noise
model's fidelity to a device does not determine whether it ranks decoders in the same order as the device. This reframes several questions that have been considered settled.

We first consider the practical use of a noise model. If a model's value in
decoder selection comes from its circuit-level structure rather than its
calibration, then the effort spent fitting a model tightly to a device may be
misdirected for this purpose, and a more useful target is a model that estimates reliability. An estimator that carries per-edge
uncertainty rather than a point estimate lets a decoder discount the edges that corrupt the ranking. Building and testing such an estimator is the natural next step, and it
would directly settle whether ill-conditioning, rather than code distance
itself, drives the inversion we observe at $d = 7$. This question cannot be answered with a single high-distance patch, but a device with several patches
can. A related limitation is that no model in our
ladder reproduces the basis asymmetry the device exhibits, so a noise model
faithful in rank need not be faithful in every omitted physical effect.

The second question concerns how vendor and neural decoders are validated. Our
case study gives the first independent evaluation of NVIDIA's Ising pre-decoder
on hardware, and shows that a decoder may show synthetic success
yet hold no advantage at code distances that exist today. As more such
datasets are released, we expect re-evaluating published decoders against them should
become routine. We built this study to clarify this, including the released pipeline runs from raw Willow ingest through every
noise-model rung, decoder, and score, so a new decoder, a new noise model, or a
new device can be compared by decoding its own shots.

\appendices
\section{Validation Gate}
\label{subsec:gate}

\begin{figure}[t]
\centering
\includegraphics[width=\columnwidth]{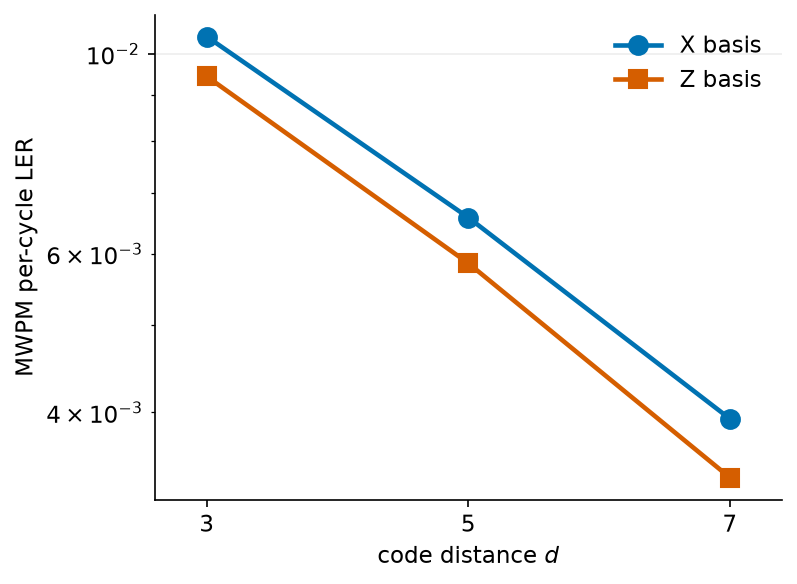}
\caption{Per-cycle MWPM logical error rate versus code distance on Willow,
both bases.}
\label{fig:below_threshold}
\end{figure}

Three checks gate every comparison in this paper, and all three pass. Decoders
run on our derived detection events reproduce the LER obtained from the shipped
events. Uncorrelated MWPM never outperforms the shipped correlated-matching
reference at any round count. Per-cycle MWPM LER falls monotonically with code
distance, the below-threshold signature (Fig.~\ref{fig:below_threshold}).

Each distance step suppresses mean per-cycle MWPM LER by a factor of $1.6$ to
$1.7$, the error-suppression factor $\Lambda$ (Willow rows of
Table~\ref{tab:tab_ladder_ler}). $\Lambda$ rises with decoder strength, from
$1.6$--$1.7$ for uncorrelated matching to $1.9$ for BeliefMatching and $2.1$
for Tesseract, which recovers the $\Lambda\approx2.14$ reported for the
correlated and neural decoders.

\section{Experimental Details}
\label{app:expdetails}

Every decode ran on 96-core CPU nodes. The released harness pins every library version.

Latency is the wall-clock time of one batched decode over a configuration's
shots, divided by the shot count. Parsing the detector error model, building
the matching graph, and loading network weights fall outside the timed call,
so the reported figure is a throughput.

MWPM and MWPM-RL run all fifteen round counts. The rest run
$r \in \{1, 10, 13, 30, 50, 70\}$ on Willow and $r \in \{1, 10, 13, 30\}$ on
the synthetic rungs. The two Ising models skip $r = 1$, which is below their
two-round minimum, so their 140 Willow configurations are the 280 evaluations
of Section~\ref{sec:ising}. Ten cells deviate from
Table~\ref{tab:run_matrix}. Two
BP+OSD cells at $d = 7$, $r = 70$ on Willow are missing, both outside the
comparison window. The other eight are BP+LSD cells at $d = 7$ on the uniform
rung, which ran at a quarter of the tabulated shots.

The sweep runs at $r = 30$ and raises a fitted parameter $\hat{p}$ to
$\min(0.5, \max(2\hat{p}, \hat{p} + 10^{-3}))$. The additive floor keeps the
perturbation real for parameters the fit left at zero. Companion arms repeat
the sweep at $r = 10$ and $r = 50$, and with a $0.5\times$ underestimate in
place of the overestimate.

Each bootstrap draw resamples every patch's fit shots with replacement, refits
the model, and re-evaluates the panel on syndromes drawn from the refit. The
40 draws are compared against the point-estimate fit, and the fraction in
which a pair of decoders changes order is that pair's rank-flip probability.

\section{Lattice Orientation Probe}
\label{app:isingvalid}

The mapping of Section~\ref{subsec:adapter} places Willow's XZZX stabilizers on
the CSS grid in one of four orientations. We re-decode the $d = 7$ patch at
$r = 70$ under all four (Table~\ref{tab:tab_rotation_probe}). XV is the only
orientation that decodes both bases. Each $Z$-type orientation fails one basis
outright, at a per-cycle rate near $4\times10^{-2}$, which accumulates to
chance over 70 rounds. ZH lowers the $X$-basis rate 34\% below XV's, to twice
MWPM's, and leaves the $Z$ basis at chance. Ising's error rate exceeds MWPM's
under all four orientations, so orientation does not explain the result of
Section~\ref{sec:ising}.

\begin{table}[t]
\centering
\caption{Ising per-cycle logical error rate at $d = 7$ and $r = 70$ under the
four CSS lattice orientations, both model sizes, against the MWPM reference.}
\label{tab:tab_rotation_probe}
\begin{tabular}{lrrrr}
\toprule
Memory basis & Rotation & Ising-fast & Ising-acc & MWPM ref \\
\midrule
X & XV & 0.0148 & 0.0145 & ~0.0049 \\
X & XH & 0.0175 & 0.0184 &  \\
X & ZV & 0.0390 & 0.0330 &  \\
X & ZH & 0.0097 & 0.0088 &  \\
Z & XV & 0.0119 & 0.0116 & ~0.0038 \\
Z & XH & 0.0132 & 0.0146 &  \\
Z & ZV & 0.0178 & 0.0212 &  \\
Z & ZH & 0.0450 & 0.0290 &  \\
\bottomrule
\end{tabular}
\end{table}

\section{Full Noise Model Results}
Tables~\ref{tab:tau_ladder} and~\ref{tab:tab_ladder_ler} list the rank
agreement and the mean per-cycle logical error rate of every decoder on every
noise model and on Willow, by code distance.
\begin{table}[H]
\centering
\setlength{\tabcolsep}{3pt}
\caption{Kendall's $\tau$ between each synthetic rung's decoder ranking and the
Willow ranking, per code distance, over the rounds 2--30 window. $p$-values test against a random ordering.}
\label{tab:tau_ladder}
\begin{tabular}{lccc}
\toprule
Noise model & $d=3$ & $d=5$ & $d=7$ \\
\midrule
uniform & 0.619 (p=0.069) & 0.714 (p=0.030) & 0.619 (p=0.069) \\
SI1000 & 0.905 (p=0.003) & 0.905 (p=0.003) & 1.000 (p$<$0.001) \\
25-param fit & 0.905 (p=0.003) & 0.905 (p=0.003) & 0.810 (p=0.011) \\
syndrome DEM & 1.000 (p$<$0.001) & 0.905 (p=0.003) & 0.905 (p=0.003) \\
\bottomrule
\end{tabular}
\end{table}
\begin{table}[t]
\centering
\caption{Full fidelity-ladder mean per-cycle LER matrix (rounds 2-30 window), decoder $\times$ code distance, one block per rung.}
\label{tab:tab_ladder_ler}
\begin{tabular}{lrrrr}
\toprule
Rung & Decoder & d=3 & d=5 & d=7 \\
\midrule
uniform & Tesseract & 0.00590 & 0.00383 & 0.00236 \\
 & BeliefMatch & 0.00609 & 0.00481 & 0.00365 \\
 & MWPM & 0.00772 & 0.00831 & 0.00780 \\
 & BP+OSD & 0.01418 & 0.03380 & 0.13394 \\
 & BP+LSD & 0.03697 & 0.05456 & 0.14913 \\
 & Ising-fast & 0.01962 & 0.03518 & 0.05277 \\
 & Ising-acc & 0.01940 & 0.03305 & 0.05067 \\
SI1000 & Tesseract & 0.00089 & 0.00012 & 0.00001 \\
 & BeliefMatch & 0.00105 & 0.00016 & 0.00003 \\
 & MWPM & 0.00136 & 0.00028 & 0.00005 \\
 & BP+OSD & 0.00172 & 0.00025 & 0.00006 \\
 & BP+LSD & 0.00279 & 0.00063 & 0.00012 \\
 & Ising-fast & 0.00474 & 0.00695 & 0.01160 \\
 & Ising-acc & 0.00482 & 0.00675 & 0.01146 \\
25-param fit & Tesseract & 0.00438 & 0.00168 & 0.00069 \\
 & BeliefMatch & 0.00505 & 0.00225 & 0.00093 \\
 & MWPM & 0.01352 & 0.00525 & 0.00280 \\
 & BP+OSD & 0.00601 & 0.00376 & 0.00177 \\
 & BP+LSD & 0.00896 & 0.00574 & 0.00276 \\
 & Ising-fast & 0.01403 & 0.02010 & 0.02959 \\
 & Ising-acc & 0.01415 & 0.01894 & 0.02883 \\
syndrome DEM & Tesseract & 0.00393 & 0.00156 & 0.00061 \\
 & BeliefMatch & 0.00462 & 0.00223 & 0.00109 \\
 & MWPM & 0.00681 & 0.00440 & 0.00261 \\
 & BP+OSD & 0.00581 & 0.00360 & 0.00213 \\
 & BP+LSD & 0.00926 & 0.00542 & 0.00312 \\
 & Ising-fast & 0.01367 & 0.02052 & 0.02860 \\
 & Ising-acc & 0.01399 & 0.01944 & 0.02753 \\
Willow & Tesseract & 0.00738 & 0.00352 & 0.00167 \\
 & BeliefMatch & 0.00815 & 0.00421 & 0.00217 \\
 & MWPM & 0.00995 & 0.00622 & 0.00366 \\
 & MWPM-RL & 0.00952 & 0.00601 & 0.00355 \\
 & BP+OSD & 0.00936 & 0.00659 & 0.00432 \\
 & BP+LSD & 0.01309 & 0.00959 & 0.00591 \\
 & Ising-fast & 0.01466 & 0.02003 & 0.02691 \\
 & Ising-acc & 0.01496 & 0.01907 & 0.02599 \\
\bottomrule
\end{tabular}
\end{table}

\end{document}